# ENERGY BALANCING THROUGH CLUSTER HEAD SELECTION USING K-THEOREM IN HOMOGENEOUS WIRELESS SENSOR NETWORKS

Muhammad Imran, Asfandyar khan, Azween B . Abdullah
Department of Computer & Information Sciences
Universiti Technologi PETRONAS
Bandar Seri Iskandar, 31750 Tronoh, Perak, Malaysia.
*cmimran81@yahoo.com*, *asfand43@yahoo.com*, *azweenabdullah@petronas.com.my*

**Abstract-** *The objective of this paper is to increase life time of homogeneous wireless sensor networks (WSNs) through minimizing long range communication and energy balancing. Sensor nodes are resource constrained particularly with limited energy that is difficult o r impossible to replenish. LEACH (Low Energy Adaptive Clustering Hierarchy) is most well -known cluster based architecture for WSN that aims to evenly dissipate energy among all sensor nodes. In cluster based architecture, the role of cluster head is very c rucial for the successful operation of WSN because once the cluster head becomes non functional, the whole cluster becomes dysfunctional. We have proposed a modified cluster based WSN architecture by introducing a coordinator node (CN) that is rich in term s of resources. This CN take up the responsibility of transmitting data to the base station over longer distances from cluster heads. We have proposed a cluster head selection algorithm based on K - theorem and other parameters i.e. residual energy, distance to coordinator node, reliability and degree of mobility. The K -theorem is used to select candidate cluster heads based on bunch of sensor nodes in a cluster. We believe that the proposed architecture and algorithm achieves higher energy efficiency through minimizing communication and energy balancing. The proposed architecture is more scalable and proposed algorithm is robust against even/uneven node deployment and node mobility.*

## 1. Introduction

A Wireless sensor network (WSN) is composed of large number o f tiny low powered sensor nodes and one or multiple base stations (sinks). These tiny sensor nodes consist of sensing, data processing and communication components. The sensor nodes sense, measure and collect ambient environment conditions , use their processing abilities to carry out simple computations and send partially processed sensed data to a base station either directly or through a gateway. The gateway can perform fusion of the sensed data in order to filter out erroneous data and anomalies and to draw conclusions from the reported data over a period of time. A comprehensive overview of wireless sensor networks and their broad range of applications can be found in [1 -3].

The sensor nodes in a wireless sensor networks (WSN) are resource constrained i.e. limited energy and computation power (processor and memory), short communication range and low bandwith. A sensor node operates on limited battery power and it is very difficult or impossible to recharge or replace it. When it is depleted of energy, it will die and disconnect from the network which significantly effect network performance. Life of a sensor node determine the lifetime of the network. Maximizing lifetime of the network involves energy conservation and harvesting. Energy is conserved th rough optimizing communication and minimizing energy usage. [1 -3].

Energy efficiency is one of the most important issue in such resource constrained environment, therefore designing energy aware routing protocols significantly extend lifetime of sensor network. Wireless sensor networks may consist of hundreds or even thousands of sensor nodes, so designing and successful operation of such large size network requires scalable architecture.

Routing in wireless sensor networks is very challenging task due to their unique characteristics, dynamic nature, architecture and design issues and resource constraints. There are various routing protocols proposed in literature but hierarchical or cluster based protocols are energy efficient, scalable and tends to pro long network lifetime; they are summarized in [4, 5]. A generic cluster -based WSN architecture is presented in figure 1. LEACH [6, 7, and 8] is one of the most popular and promising cluster based protocol that have been widely proposed in WSNs. It uses mul tihop communication when a node is unable to reach base station. The cluster heads performs the additional responsibilities like collecting data from all the sensor nodes in a cluster and transmits it to base station over long distances. The role of cluster head is rotated after each round among all the nodes present in a cluster to balance the energy level. It is assumed that each sensor node has a long range communication and is able to reach CH directly and thereafter BS. Various improvements can be made to make it more energy efficient and scalable.

In this paper, we have proposed a modified cluster based WSN architecture, where we have introduced an idea of coordinator node (s) that is rich in terms of resources. We have proposed an algorithm for effici ent selection of cluster heads based on different parameters. We believe that through our modified architecture, lifetime of WSN is increased through energy balancing and it also increases scalability of network. We also believe that our proposed algorithm is robust against even or uneven node deployment, node mobility. It results in an efficient cluster head selection to evenly distribute the energy dissipation and hence increase of network lifetime.

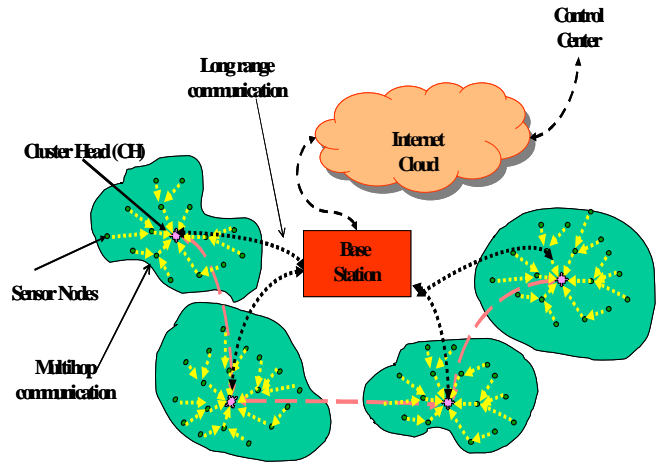

Figure 1.Cluster-based wireless sensor network architecture [19].

The rest of the paper is organized as follows. Section II presents critical review of related work. In section III, we have described the problem statement. In section IV, we have presented our proposed architecture and algori thm for energy balancing and cluster head selection. Section V presents the analytical discussion of proposed architecture and algorithm. We have concluded our work in section VI.

## 2. Related work

Energy efficient communication is a matter of survival for w ireless sensor networks. Because of the reason, most of the research is focused towards energy efficient communication, energy conservation, maximizing network lifetime and energy harvesting. Various routing strategies have been proposed but cluster -based routing protocols are dominant and considered energy efficient and scalable. In [9], a framework to select the optimal probability with which a node should become a cluster head in order to minimize network energy consumption has been proposed. They have also proposed a sleep-wakeup based MAC protocol instead of TDMA for LEACH architecture.

L. Ying et al [10] proposed an energy adaptive cluster head selection algorithm based on node residual energy and energy required for transmission. In [11], energy effi cient cluster ID based routing scheme have been proposed, where, uneven load in network is minimized by cluster size adaptation. The current CH keeps the information of node with maximum remaining energy in its cluster. It ensures the location of CH approx imately at center of cluster. In case of uneven deployment it is not necessary that most of the nodes are in center of cluster. So the distance between CH and sensor nodes will be increased. Another protocol presented in [12], consider distance and remaining energy of a node. Base station determines and select appropriate cluster head based on above two parameters. Y. Yin et al [13] proposed a centralized CH selection mechanism based on energy, mobility and distance to cluster centroid. They assume that se nsor nodes are location-aware and there is a single-hop communication with in a cluster.

Most of the protocols proposed so far only consider residual energy for cluster head selection but there are other metrics like distance to aggregation point, node re liability, mobility [20] etc that are also very crucial in order to maximize network life time. Appropriate energy balancing protocols are required to increase scalability, reduce energy consumption and enhance the network lifetime of the network.

## 3. Problem statement

The sensor nodes in a wireless sensor network (WSN) are resource constrained particularly with limited energy that is difficult or impossible to replenish. Transmission is the main energy guzzler and it further increases with the distance, as energy consumption is directly proportional to the square of the distance between the nodes. So in order to conserve energy and increase network lifetime, we had to minimize energy dissipation and optimize communication. It is possible through efficient ro uting schemes. Scalability is another major issue, as they may contain hundreds and thousands of sensor nodes. Both of the above issues have been addressed efficiently by cluster-based architectures and algorithms particularly LEACH. There are few areas in LEACH that can be improved to make it more energy efficient and scalable.

In LEACH, the role of cluster head is very crucial for the successful operation of WSN because once the cluster head becomes the non functional, the whole cluster becomes dysfunctional. In LEACH, CHs are selected at random; there is good chance that a node with very low energy, vulnerable or isolated node in a cluster gets selected as a CH. In each round new cluster head is elected and new routing table is formed that if avoided, significant amount of energy can be conserved [11, 14]. Cluster heads are over-loaded with long range transmissions to the base stations [11] or sink and with additional processing responsibility of data aggregation/fusion before transmission [9]. Due to these responsibilities, CH nodes drained through their energy quickly. There is an assumption that each sensor node has long communication range to reach cluster head and thereafter BS. It is unrealistic and inefficient particularly in case of homogeneous WSN that are deployed in large regions, where sensor nodes have limited transmission range and CHs are regular sensor nodes. The CHs near the base station drained through their energy quickly because of multihop inter-cluster communication. This leads to energy imbalance among the clusters.

The proposed algorithm assumes that
- Clusters are already formed during setup phase using LEACH..
- BS is assumed to be stationary and all coordinator nodes are able to reach BS either direct or multihop routing.
- Nodes have power control features so as to adjust their transmission power to minimum level required for successful transmission based on RSSI (Received Signal Strength Indicator) [10, 15].

## 4. Proposed work

Our proposed work consists of two parts. First, we have proposed modification to the cluster based WSN architecture by introducing Coordinator Node (CN). The detailed working of modified architecture is described in following section. Second, we have proposed an algorithm for cluster head selection in order to achieve energy balancing in a cluster based on LEACH protocol. The detailed algorithm is described in section B.

### 4.1. Modified cluster based homogeneous WSN architecture

In our modified architecture, we have proposed a new Coordinator Node (CN) that is rich in resources compared to other sensor nodes. It has more computation resources (more processing power and memory) and energy. It has longer transmission range to reach base station. The coordinator node is reachable to all the nodes in all clusters. If it is not reachable, it is recommended to add another coordinator node. It can be a special node or devices like laptop or PDA can be used for the purpose. Our modified architecture tends to distribute the load of CH to the CN.

The coordinator node is responsible for cluster head selection. The purpose of introducing CN is to closely monitor the operation of sensor nodes in a cluster and command them for specific operations.

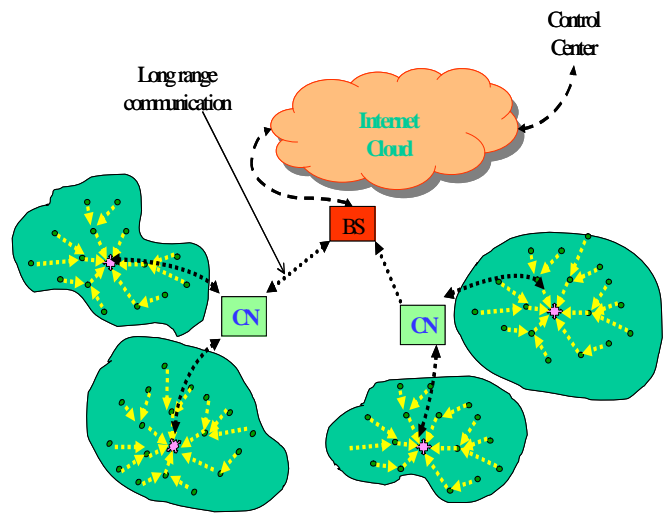

Figure 2. A modified cluster-based architecture for wireless sensor networks where sensor nodes send the sensed information to the cluster-heads through multihop routing. CHs aggregate the received information and transmit it to the Coordinator Node (CN), which then forward it to the base station.

All the sensor nodes deployed in the sensor field are considered as homogeneous i.e. same capability in terms of computation and communication. The nodes may be mobile that can move around in a sensor field. These sensor nodes sense, measure and collect ambient environment conditions and transform them into electric signal. These nodes use their processing abilities to carry out simple computations and send partially processed sensed data to corresponding cluster head via the radio transmitter.

### 4.2. Algorithm description

Our proposed algorithm tends to balance energy in a cluster through energy efficient reliable cluster head selection. The objective of proposed algorithm is to improve the cluster head selection mechanism of LEACH. The operation of LEACH is divided into rounds. Each round consists of a setup phase and steady-state phase. In setup phase, coordinator node selects the cluster head for each cluster. In steady-state phase, sensor nodes sense the environment and transmit the sensed data to the corresponding CH for further onward transmission to the CN and BS.

**Setup phase:**
Our work assumed that clusters are already formed in setup phase and coordinator node is aware of the cluster formation and information. Our proposed algorithm for cluster head selection consists of following steps:

**Step-I:** The CN set the value of k for the current round for each cluster based on density of nodes in a cluster. It broadcast the value of k to each corresponding cluster i.e. $k_i$. The value of k determines the k number of nearest neighbor nodes.

**Step-II:** All the sensor nodes send their k number of nearest neighbors (based on distance) to the CN. The distance to the node can be calculated based on *received signal strength indicator* (RSSI).

**Step-III:** The coordinator node select candidate set of cluster heads i.e. $C_i$ for each cluster through K-theorem. The value of $k_i$ is always equal to the number of candidate cluster heads in a cluster i.e. $C_i$. The detailed working of K-theorem is described in section D.

**Step-IV:** The CN request candidate set of cluster heads in each cluster to send their combined rating (CR).

**Step-V:** Each candidate cluster head node calculate it's own CR based on residual energy (RE), distance to coordinator node, node reliability (R) and degree of mobility (M) and send it to CN. The phenomenon of calculating combined rating is described in equation.

**Step-VI:** The coordinator node selects a node as cluster head among candidate set of cluster heads for each cluster based on CR. The higher the CR a node has; greater the chances of being cluster head. The CN confirms each cluster about their CH.

After the CH selection, routing paths are established that is beyond the scope of this research except our research prefers multihop routing. Each CH creates a TDMA (Time Division Multiple Access) schedule for intra-cluster communication. In TDMA, one-time slot is reserved for each cluster member to send or receive its data to/from the CH.

**Steady phase:**
In steady phase, each sensor node wake up in its allocated time slot and send or receive data to/from the CH. The corresponding CH aggregates the collected data from its cluster and forwards it to the CN. The CN may apply some compression algorithm on received data from all the CHs and transmit it to the base station either directly or through multihop transmission.

At the end of each round, each CH check its residual energy and energy dissipated in previous round. The decision for reselection of cluster head is based on residual energy and energy dissipated by the cluster head in previous round. If the residual energy is twice the energy dissipated in previous round, the CH would continue for the next round. On the other hand if residual energy is twice the energy dissipated in previous round and there are significant number of nodes alive in a cluster (may be equal to value of $k_i$). It would consult to CN for CH reselection. A node can only serve as CH twice consecutively. The purpose is to avoid CH reselection after each round because the process itself requires significant amount of energy.

The nodes can exchange network setup and maintenance messages instead setup phase (if there is no need to reselect cluster head). Information about topological changes (due to dead nodes or node mobility) can be exchange during this phase.

### 4.3. Working of K-theorem

The philosophy behind the K-theorem is to select a candidate CHs based on bunch of sensor nodes in a cluster. The working of K-Theorem is quiet simple and was proposed to select optimal server location. Table II shows the working details.

The coordinator node set the value of $k_i$ for each cluster. The value of $k_i$ is relative to the node density in a cluster and ratio (i.e. r) of the cluster heads in a WSN. It is the product of the number of nodes in a cluster (i.e. $n_i$) and ratio r. The value of r can vary can vary from 0.01 to 0.99 but it should not be more than 0.50. The less the value of $k_i$ is; the more the probability of getting local optima is. The value of $k_i$ determines the $k_i$ number of best sensor nodes that can serve as cluster head. The value of $k_i$ also provides the alternative suboptimal options, so that we can select an optimal sensor node for cluster head.

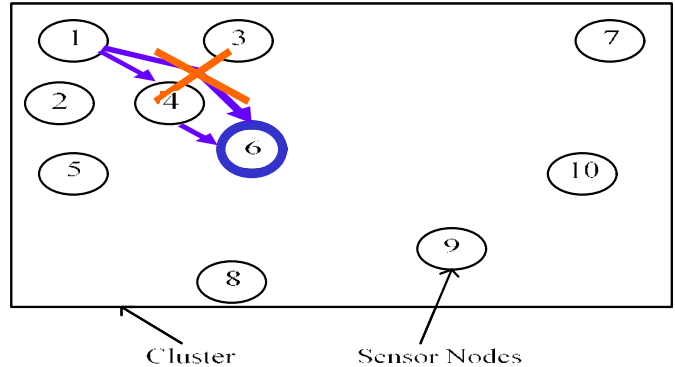

Figure 4 shows the working of K-theorem based on multihop route selection. Where each node selects it's k nearest neighbors based on received signal strength indicator (RSSI).

For each sensor node deployed in the cluster, we choose it's $k_i$ nearest neighbors based on distance. The distance between sensor nodes can be calculated through received signal strength indicator (RSSI) that is described in detail in [15] or any other localization technique [16, 17]. Multihop communication route is preferred while finding nearest neighboring nodes when distance is larger. As multihop communication is energy efficient compared to direct communication [1, 3], so choosing a neighbor having multihop connection will consume less energy. As our objective is energy efficient cluster head selection, so we will prefer multihop communication over direct communication. The phenomenon of multihop route selection is depicted in figure 4.

Then frequency of occurrence of each sensor node is calculated and list in table-I. The ordered list of sensor nodes based on their frequency is shown i.e. $S_i$. The minimum frequency required in cluster i to be the CH (i.e. $K_i$) is calculated based on weighted mean of frequencies and 1 is added for better result. Weighted mean is calculated by product of each frequency of occurrence into number of sensor nodes having that frequency. The value of $K_i$ is rounded to the nearest integer if required. The value of Sensor nodes having frequency $K_i$ or greater are identified from Table II and are candidates for cluster head i.e. $C_i$. The candidate cluster head nodes would always be equals to value of $k_i$ i.e. 3 in this case.

Table I. List of nodes with their K-nearest neighbors and their frequency of occurrence

| Node ID | $k_i = 3$ List of Terminals with its K-Nearest Neighbor | Frequency of Occurrence |
|---|---|---|
| 1 | 1) 2, 3, 4 | 3 |
| 2 | 2) 1, 4, 5 | 4 |
| 3 | 3) 1, 4, 6 | 5 |
| 4 | 4) 2, 3, 6 | 6 |
| 5 | 5) 2, 4, 6 | 4 |
| 6 | 6) 3, 4, 5 | 7 |
| 7 | 7) 3, 9, 10 | 2 |
| 8 | 8) 5, 6, 9 | 2 |
| 9 | 9) 6, 8, 10 | 4 |
| 10 | 10) 6, 7, 9 | 3 |

Sorting {$S_i$ = Ordered list of sensor nodes, where i is the frequency of occurrence}
S2 = (7, 8)           S3 = (1, 10)
S4 = (2, 5, 9)        S5 = (3)
S6 = (4)              S7 = (6)
$K_i$ = [Weighted Mean] + 1
  = [(2*2)+(3*2)+(4*3)+(5*1)+(6*1)+(7*1) / 10 ] + 1

= [(4 + 6 + 12 + 5 + 6 + 7) / 10] + 1
= [(40) / 10] + 1      ➔ 4 + 1
$K_i = 5$
So, the best nodes for candidate cluster head in cluster i are: $C_i = \{3, 4, 6\}$ when $k_i = 3$.

### 4.4. Combined Rating

The combined rating is calculated based on following criterion:

#### 4.4.1. Residual energy ($\alpha$):

The residual energy of a node preferably be greater than the approximate energy dissipated in previous round by the cluster head. Equation for residual energy of node i is described in [12].

#### 4.4.2. Distance to coordinator node ($\beta$):

The nodes having less distance from coordinator node should have higher probability to become cluster head. As energy consumption is directly proportional to the square of distance. Distance to coordinator node for node i is expressed in [12].

#### 4.4.3. Node Reliability (R):

As the role of cluster head is very crucial for successful operation of wireless sensor network. If a cluster head stops working, the whole cluster becomes dysfunctional. A candidate cluster head sensor node may fail due to lack of energy, physical damage or environmental interference. Reliability deals with continuity of service. The purpose of node reliability is to increase the trustworthiness. The reliability $R_i(t)$ of a sensor node is modeled in [18] using Poisson distribution to capture the probability of not having a failure within the time interval (0, t):

#### 4.4.4. Degree of Mobility (M):

The life time of the network is greatly influenced due to the mobility of a node. It can lead to higher topological changes and requires frequent cluster head reselection. Information about topological changes (due to dead nodes or node mobility) can be exchange during maintenance phase.

## 5. Analytical Discussion

Energy dissipated is directly proportional to the square of the distance and transmission is the major energy guzzler. In our modified cluster-based architecture, the idea is to distribute the load of long range transmission from CH to the CN. This will conserve energy at CH and ultimately lead to increase the life time of wireless sensor network. The problem, where CHs near the BS were drained through their energy quickly, because of inter-cluster multihop communication is also addressed through modified architecture. There is no need to have inter-cluster multihop communication because each cluster head is able to directly reach the CN. In this way, the CHs near the BS or CN does not have to take extra responsibility of transmitting other clusters information. They just have to transmit their own data. This will lead to energy balancing among the clusters that was previously a problem. The modified architecture can be more scalable as we just need to add extra CN if network density or distance is high.

The proposed algorithm tends to improve CH selection based K-theorem and other four crucial parameters that contribute towards network lifetime. The benefit of using K-theorem is that it minimizes the communication and reduces the long range intra-cluster communication. It provides alternative candidate CHs so that additional criterion may be applied to select the best among the bunch of sensor nodes from a cluster. The other four parameters further improves the CH selection that have higher residual energy, less distance to reach CN, reliable node and with less degree of mobility.

There are few issues that still need to be resolved e.g. to find out the best placement of coordinator node, viability of a sensor node to become coordinator node in special circumstances where node density is very low.

## 6. Conclusion:

In this paper, we have modified the architecture of cluster based wireless sensor networks by proposing a coordinator node. This will not only increase the life time of the network but scalability of the network is also increased. It also eliminates the need of inter-cluster communication to reach the BS. We have propose an algorithm for cluster head selection based on K-Theorem and four parameters i.e. residual energy, distance to the coordinator node, reliability and degree of mobility. The K-theorem selects candidate cluster heads based on bunch of sensor nodes within a cluster. The K-theorem is robust against sensor nodes deployment in a cluster, whether the deployment is structured or unstructured, even or uneven, nodes are stationary or mobile. We believe that this will not only minimize the communication cost but will also increase the reliability of the network.

## Acknowledgement

We are thankful to all of our colleagues including Mr. Sajjad Haider, who have been very helpful in devising K-theorem. We are also thankful to our friends for their useful comments and suggestions.

## References

[1] I. F. Akyildiz, W. Su, Y. Sankarasubramaniam, E. Cayirci, "Wireless sensor networks: a survey", Computer Networks: The International Journal of Computer and Telecommunications Networking, v.38 n.4, p.393-422, 15 March 2002.

[2] J. Yick, B. Mukherjee, D. Ghosal, "Wireless sensor network survey", Computer Networks: The International Journal of Computer and Telecommunications Networking, Accepted 7 April 2008.

[3] M. Younis, K. Akkaya, M. Eltoweissy, A. Wadaa, "On Handling QoS Traffic in Wireless Sensor Networks", Proceedings of the Proceedings of the 37th Annual Hawaii International Conference on System Sciences (HICSS'04) - Track 9, p.90292.1, January 05-08, 2004.

[4] K. Akkaya, M. F. Younis: A survey on routing protocols for wireless sensor networks. Ad Hoc Networks 3(3): 325-349 (2005)

[5] A. A. Abbasi, M. F. Younis: A survey on clustering algorithms for wireless sensor networks. Computer Communications 30(14-15): 2826-2841 (2007)

[6] W. Heinzelman, A. Chandrakasan and H. Balakrishnan, "Energy-efficient communication protocols for wireless microsensor networks" Proceedings of the 33$^{rd}$ International Conference on systems Sciences, Hawaii, Jan. 2000.

[7]. W. Heinzelman, `Application-Specific Protocol Architectures for Wireless Networks," Ph.D. Dissertation, Massachusetts Institute of Technology, June 2000.

[8]. W. R. Heinzelman, A. Chandrakasan, and H. Balakrishnan, "An Application-Specific Protocol Architecture for Wireless Microsensor Networks", IEEE transactions on wireless communications, vol. 1, no. 4, October 2002.

[9]. H. Yang, B. Sikdar, "Optimal Cluster Head Selection in the LEACH Architecture", In the proceedings of the 26$^{th}$ IEEE International Performance and Communications Conference, IPCCC2007, April 11-13, 2007, New Orleans, Louisiana, USA.

[10]. L. Ying, Y. Haibin, "Energy Adaptive Cluster-Head Selection for Wireless Sensor Networks", In Proceedings of the 6$^{th}$ International Conference on Parallel and Distributed Computing, Applications and Technologies (PDCAT'05), pp. 634--638, 2005.

[11]. I. Ahmed, M. Peng, W. Wang, "A Unified Energy Efficient Cluster ID based Routing Scheme for Wireless Sensor Networks-A more Realistic Analysis", Proc of IEEE Third International conference on Networking and Services (ICNS'07), p-86, 2007.

[12]. P. Tillapart, S. Thammarojsakul, T. Thumthawatworn, P. Santiprabhob, "An Approach to Hybrid Clustering and Routing in Wireless Sensor Networks", Proc of IEEE Aerospace Conference, 2005.

[13]. Y. Yin, J. Shi, Y. Li, P. Zhang, "Cluster-Head Selection using Analytical Heierarchy Process for Wireless Sensor Networks", The 17th Annual IEEE International Symposium on Personal, Indoor and Mobile Radio Communications (PIMRC'06), China.

[14]. H. Hsu, Q. Liang, "An Energy-Efficient Protocol For Wireless Sensor Networks", In proceedings of 62$^{nd}$ IEEE Vehicular Technology Conference, 2005, vol.4, pp-2321-2325.

[15]. A. A. Minhas, "Power Aware Routing Protocols for Wireless ad hoc Sensor Networks", Ph. D Thesis, Graz University of Technology, Graz, Austria, March, 2007.

[16] L. Hu, D. Evans, Localization for mobile sensor networks, ACM International Conference on Mobile Computing and Networking (MobiCom 2004), 2004.

[17] J. Hightower, G. Borriello, Location systems for ubiquitous computing, IEEE Computer 34 (8) (2001) 57–66.

[18] G. Hoblos, M. Staroswiecki, A. Aitouche, Optimal design of fault tolerant sensor networks, IEEE International Conference on Control Applications, Anchorage, AK, September 2000, pp. 467–472.

[19]. U. B. Desai, B. N. Jain, S. N. Merchant, "Wireless Sensor Networks: Where do We Go?", www.ee.iitb.ac.in/spann.

[20]. G. Ahmed, N. M. Khan, R. Ramer, "Cluster Head Selection Using Evolutionary Computing in Wireless Sensor Networks", Progress In Electromagnetics Research Symposium, Hangzhou, China, March 24-28, 2008.